# Folding@home: achievements from over twenty years of citizen science herald the exascale era

Vincent A. Voelz [1], Vijay S. Pande[2], and Gregory R. Bowman [3]

[1] Department of Chemistry, Temple University, Philadelphia, PA 19122
[2] Andreessen Horowitz, Menlo Park, CA  94025
[3] Departments of Biochemistry & Biophysics and of Bioengineering, University of Pennsylvania, Philadelphia, PA 19104

Corresponding email: grbowman@seas.upenn.edu

## Abstract

Simulations of biomolecules have enormous potential to inform our understanding of biology but require extremely demanding calculations. For over twenty years, the Folding@home distributed computing project has pioneered a massively parallel approach to biomolecular simulation, harnessing the resources of citizen scientists across the globe. Here, we summarize the scientific and technical advances this perspective has enabled. As the project's name implies, the early years of Folding@home focused on driving advances in our understanding of protein folding by developing statistical methods for capturing long-timescale processes and facilitating insight into complex dynamical processes. Success laid a foundation for broadening the scope of Folding@home to address other functionally relevant conformational changes, such as receptor signaling, enzyme dynamics, and ligand binding. Continued algorithmic advances, hardware developments such as GPU-based computing, and the growing scale of Folding@home have enabled the project to focus on new areas where massively parallel sampling can be impactful.  While previous work sought to expand toward larger proteins with slower conformational changes, new work focuses on large-scale comparative studies of different protein sequences and chemical compounds to better understand biology and inform the development of small molecule drugs. Progress on these fronts enabled the community to pivot quickly in response to the COVID-19 pandemic, expanding to become the world's first exascale computer and deploying this massive resource to provide insight into the inner workings of the SARS-CoV-2 virus and aid the development of new antivirals. This success provides a glimpse of what's to come as exascale supercomputers come online, and Folding@home continues its work.

## Introduction

Atomically-detailed computer simulations of protein dynamics have the potential to provide insight into the mechanisms of biological processes that would be impossible to obtain by experiment alone. Molecular simulation can advance our understanding of biology, and has already become an indispensable tool in drug design and protein engineering. However, these simulations are extremely computationally demanding, prompting myriad efforts to develop clever algorithms and new hardware to accelerate these calculations.

For over twenty years, the Folding@home distributed computing project has tackled this challenge by pooling the computer power of volunteers around the world to build a planetary-scale virtual supercomputer.[1] By enabling anyone with a computer and an Internet connection to become a citizen scientist by helping run simulations on their personal computer(s), the project has achieved numerous computing records (e.g. first petascale and exascale computing platform[2]) and simulation records (e.g. first simulations of millisecond timescale processes[3]). Moreover, it has enabled scientific advances in topics ranging from our understanding of fundamental processes like protein folding[4-6] to new opportunities to treat diseases like viral infections, cancer, and neurodegeneration.

Folding@home has been a community effort from the beginning, and shows the great power of engaging the public in the scientific process. Many volunteers participate out of scientific curiosity or a personal connection to one of the diseases being addressed. Other volunteers are gamers and computing enthusiasts, attracted by the technical challenges and gamification of participation, which gives users the chance to earn points proportional to their contributions. To engage users, the scientific team invests significant effort in social media and outreach to explain the relevant scientific principles, the questions addressed, and the new insights obtained. A community-driven ecosystem has developed around the project, providing everything from technical support to scientific discussions. Many volunteers have gained a deeper appreciation for (and understanding of) science through their participation, with some going so far as to pursue STEM careers.

Recently, work on COVID-19 drove unprecedented growth in the project, creating a computing resource orders of magnitude larger than the largest supercomputers and enabling rapid progress on the global health threat that the pandemic presented.[2] This development, driven by a self-organizing community of Folding@home participants, exemplifies the collective ability of citizen scientists to help tackle critical scientific problems.

Here, we reflect on what Folding@home has accomplished and the implications for the future of computational biophysics as exascale computing becomes more accessible. Such machines will enable researchers to go big in every possible dimension. Much will be learned from simulating larger systems. However, we propose there is also much to learn from simulating many different conditions in parallel and comparing the results. For example, comparing the dynamics of multiple variants will increase our understanding of protein function and comparing the binding of many compounds will accelerate the discovery of new drugs. We focus on proteins in this review, but note that Folding@home is equally applicable to other molecular systems, and has been applied fruitfully to molecules like RNA,[7,8] lipids,[9] and carbon nanotubes.[10]

## The inspiration for Folding@home

Folding@home was originally conceived at Stanford University by Vijay Pande to elucidate fundamental mechanisms of protein folding, the process by which proteins self-assemble into functioning molecular machines.[1] At the time, there was great excitement at the prospect that computer simulations could provide an atomically-detailed picture of how proteins fold that could not be achieved via any conceived experimental technique. However, reaching the relevant timescales even for very small model systems was far beyond reach of any existing supercomputer, much less commodity hardware.

The Napster platform for sharing music (as .mp3 files) provided a source of inspiration. Storing a large library of music at a central location and distributing it over the internet would have been a costly endeavor. However, Napster's peer-to-peer approach to file sharing allowed a large community of users to come together to create a powerful file sharing service. Could the protein folding problem also be broken down into independent pieces that could be performed by a large, distributed community?

This inspiration spurred the Pande lab to develop an algorithm that broke the protein folding problem down into a large number of completely independent simulations.[11] The algorithm made the simplifying assumption that protein folding can be modeled as a two state process with a single unfolded state separated by a large free energy barrier from a single folded state. A single simulation started from the folded state would make repeated attempts to cross the barrier into the folded state. Given enough simulation time, it would make many transitions back and forth between the folded and unfolded state, and one could infer the folding rate and the probabilities of the two states from the simulation. The key insight was that these attempts to cross the barrier between the folded and unfolded states did not need to be made serially. One could instead run many independent simulations started from the unfolded state, each of which would make independent attempts to transition to the folded state. Quantities such as the folding rate could then be inferred from the relative number of simulations that successfully reached the folded state compared to the total number of simulations. Roughly speaking, capturing a microsecond timescale folding process didn't require a single multi-microsecond long simulation, one simply needed microseconds of total simulations time divided amongst different simulations. Of course, the model would work well for two-state systems and break down for more complex ones (Fig. 1). However, two-state systems were a great place to start given they are some of the smallest and fastest folding proteins, where simulations and experiments can first meet.

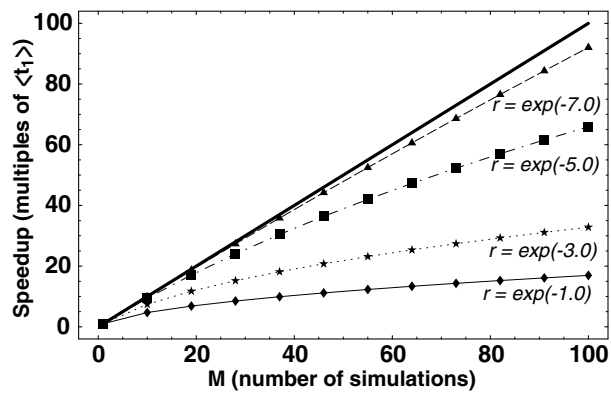

Fig. 1. Speedup versus number of simulations for a three-state (two-barrier) system. The key conclusion is that the more two-state a system is (i.e. the more one barrier is much larger than the other), the closer one comes to linear scaling, where running $M$ simulations of length $t_1$ is equivalent to running one simulation of length $M \times t_1$. Shown are plots for a range of $r = k_{12}/k_{23}$, where $k_{12}$ is the rate constant of the slow barrier crossing, and $k_{23}$ is the rate constant of the fast barrier crossing. Reproduced from Shirts and Pande.[11]

In the year 2000, the Pande lab announced the Folding@home distributed computing project to enable anyone with a computer and an internet connection to contribute their personal computing power to run large numbers of completely independent simulations of protein folding, which would be aggregated and used to gain insight by the scientific team. Google took note and added a button to the toolbar they distributed for searching the web that allowed users to volunteer their computers to contribute to Folding@home. Many people opted to participate in Folding@home by letting the software use their computing power when their machines were otherwise idle. Folding@home quickly began setting computing records. For example, it holds the Guinness world record for being the first petascale computer, capable of performing a million billion operations per second. More recently, Folding@home became the first exascale computer, capable of performing a billion billion operations per second.[2]

## Insights into protein folding

Folding@home enabled a number of early successes in capturing the atomic details of the folding of small model systems.[4] For example, Zagrovic captured the folding of a β-hairpin.[12] Snow et al. then used Folding@home to capture the folding of a small mini protein, called BBA5.[13] They predicted the folding rate, and good agreement with the prediction was found in corresponding laser temperature jump experiments. In parallel, the Pande lab showed that simulations were sufficiently accurate to fold the villin headpiece,[14] a small fast folding protein that served as one of the primary systems where scientists could connect simulations and experiments in the coming years.

Many of the early applications of Folding@home found that even the simplest systems had complex folding dynamics with multiple intermediate states, either those known experimentally or with transient intermediates with lifetimes much longer than a single simulation, breaking the applicability of the original two state approximation regime. Together with the desire to tackle even more complex systems, this insight prompted work on the development of new methods for moving from a two-state framework for protein folding to a multi-state view. In 2004, Singhal et al. published the Pande lab's first papers on Markov state models (MSMs).[15] The core idea was to build a network model of a protein's conformational ensemble, with states (or nodes) corresponding to free energy minima where a protein tends to dwell and links representing the probability of hopping between pairs of states in a fixed time interval, called the lag time of the model. This paradigm quickly provided new insights into the folding of small proteins, like villin.[16-18]

One of the major insights stemming from work on protein folding is that the free energy landscapes of many proteins have a hub-like topology, with many parallel paths leading to the folded state.[5] This observation was first made for the 35-residue villin headpiece.[19] Given that this protein is so small and fast folding (μs timescale folding), an open question was how general this behavior was. Voelz et al. then captured the first millisecond timescale folding process and found the same hub-like topology (Fig. 2).[3] Furthermore, they observed that different folding pathways resembled different models of protein folding that had been proposed based on experiments performed on different systems. They suggested that mutations likely alter the relative probabilities of such pathways, changing the experimentally observed folding mechanism. Similar observations were later made for other proteins.[20,21]

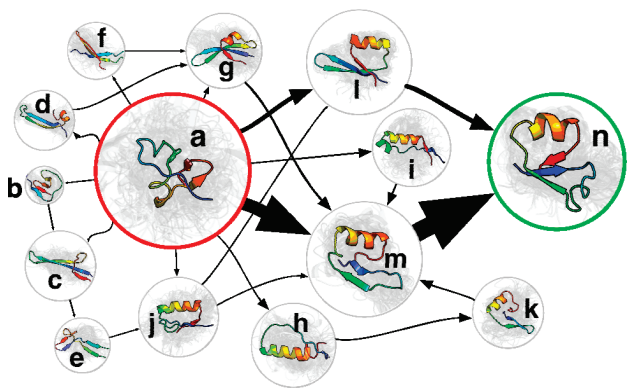

Fig. 2. The 10 highest-flux folding pathways for the 39-residue protein NTL9, which folds on a millisecond timescale. State sizes are proportional to the population, arrow widths are proportional to the flux, the colored ribbon shows a representative structure, and the gray structures convey the extent of structural diversity. Reproduced from Voelz et al.[3]

## Misfolding and neurodegeneration

While Folding@home's first priority was to understand protein folding, understanding the role of dynamics in protein function and dysfunction was always a larger goal. One of the first applications of Folding@home in this direction was to protein misfolding diseases, like Alzheimer's and Huntington's. Disordered proteins that are difficult to study using standard structural techniques are often implicated in these diseases. In principle, simulations are just as applicable to these systems as to folded proteins. Of course, adequately sampling the relevant degrees of freedom is challenging for such heterogeneous systems. Fortunately, the same principles that have enabled Folding@home to model the heterogeneous unfolded states of proteins and their folding transitions have enabled progress on disordered proteins and misfolding. For example, Kelley et al. proposed a mechanism for how N17 of the Huntingtin protein initializes dimerization and nucleates aggregation.[22] Small populations of β-sheet-rich structures were also found when studying the folding of helical proteins, suggesting such structures may be a common feature that provides an opportunity for aggregation.[20]

Simulations performed on Folding@home also provided a number of insights into Aβ and its role in Alzheimer's disease (AD). Lin et al. proposed that the 42-residue version of Aβ is more pathogenic because it stabilizes a β-hairpin that may nucleate oligomerization more than the shorter 40-residue version of Aβ.[23] Based on this insight, Novick et al. designed variants of Aβ and small molecules that inhibited aggregation.[24] Simulations were also used to understand how familial mutations associated with AD change Aβ's structural preferences and, ultimately, AD risk.[25]

More recently, Folding@home simulations helped provide unprecedented insight into apolipoprotein E (ApoE),[26] which is the strongest genetic risk factor for AD.[27] Like Aβ and other disordered proteins, ApoE has been difficult to study due to its propensity to form heterogeneous oligomers at very low concentrations. Simulations are complicated by the fact that ApoE has both well-folded and disordered regions. However, running milliseconds of simulation on Folding@home has proved capable of accounting for this spectrum of different degrees of heterogeneity, providing atomically-detailed models that are in excellent agreement with single molecule experiments and provide a basis for understanding how variants modulate AD risk and for designing new therapeutics.

## Conformational change

In the last decade, Folding@home's focus has largely shifted from protein folding to other conformational changes of importance for protein function, dysfunction, and the design of new drugs and proteins.[28-30]

One of the first focal points was understanding the relevance of fundamental concepts like induced fit and conformational selection. Induced fit posits that protein conformational changes are caused by interactions with binding partners, whereas conformational selection posits that proteins are constantly undergoing spontaneous conformational changes and that binding partners shift the relative probabilities of the different conformations that are available. Addressing this point is of practical importance because it determines what must be simulated. If conformational change dominates, then much can be learned from simulations of isolated proteins. In contrast, if induced fit dominates, the simulations must explicitly include a protein's binding partner(s) to observe functionally-relevant conformational changes.

Much evidence supports the importance of conformational selection, making simulations of individual proteins highly informative in many cases. For example, analysis of simulations of WW domains suggested that conformational selection is at play in their binding interactions.[31]

These simulations did not include any of the WW domain's binding partners, so they could not directly assess the relative importance of conformational selection and induced fit. Simulations of the Lysine-, Arginine-, Ornithine-binding (LAO) protein and one of its substrates, arginine, directly addressed the relative roles of conformational selection and induced fit, finding a significant role for conformational selection (Fig. 3).[32] Specifically, the authors found that binding occurs via a 3-state mechanism, in which the protein exists in an equilibrium between an open and partially closed state. Ligand interacts favorably with the partially closed state and can then induce total closure, which is extremely rare in the absence of ligand. Later studies of the binding of the p53 TAD peptide and its protein receptor MDM2 also suggested a mixture of conformational selection and induced fit.[33] Importantly, this work recognized that the dominant mechanism is dependent on the concentrations of the species, and found that induced fit dominates across a wide range of concentrations in this particular case. A study of cyclic peptide binding to MDM2 found that the idea of conformational selection applies just as well to foldable ligands. A mix of conformational selection and induced fit was also observed in the binding of a ligand to β-lactamase.[34] Together, these studies suggest that conformational selection plays a sufficiently significant role that much can be learned from simulations of isolated proteins without explicitly including their binding partners in the simulation.

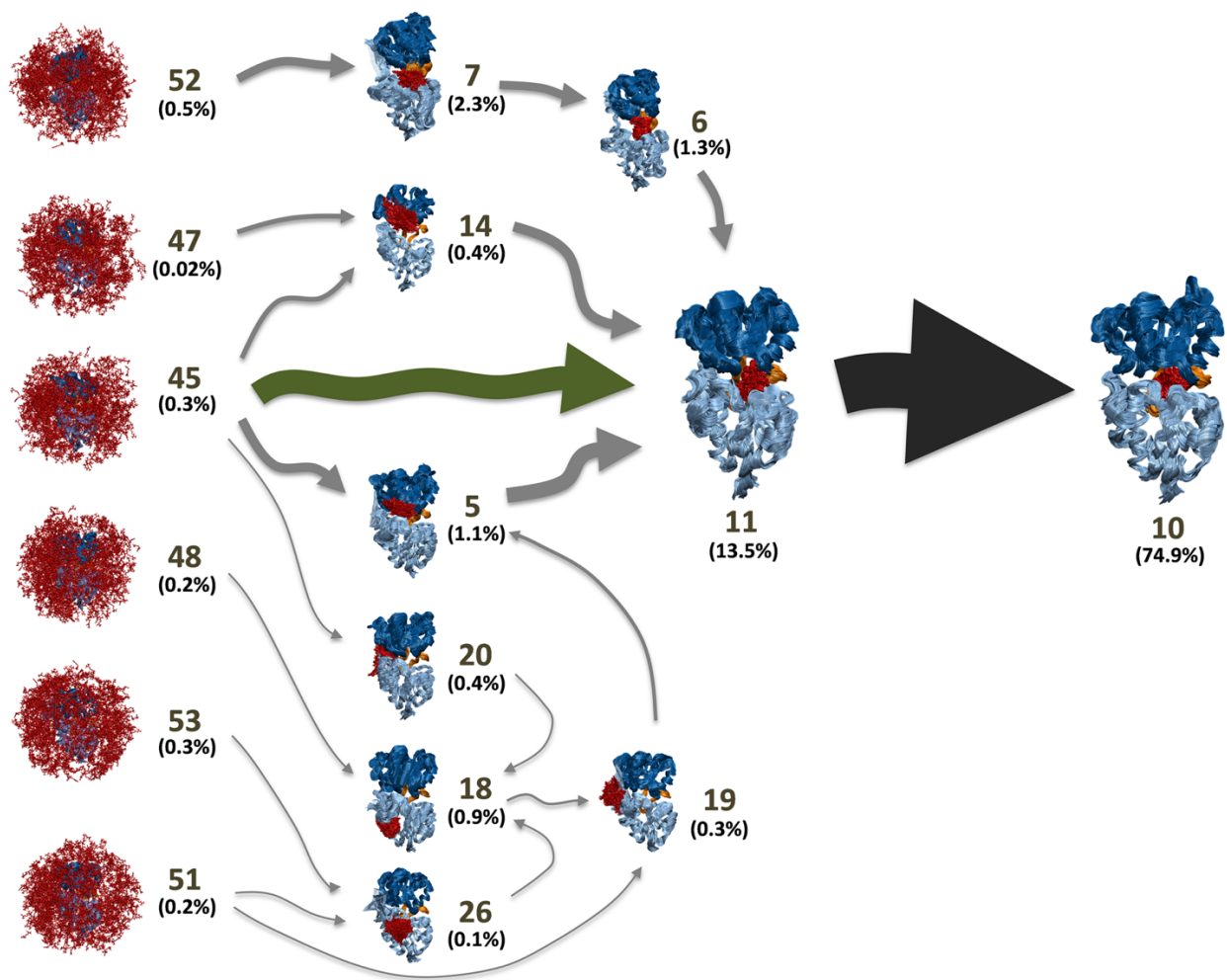

Fig. 3. The 10 highest flux pathways from the unbound states of the LAO protein to the arginine-bound state. The arrow widths are proportional to the flux, the two lobes of the LAO protein are shown in dark/light blue, and arginine is shown as red sticks. State numbers and their equilibrium populations are also shown. The conformational selection and induced fit pathways from the unbound states to the encounter complex state are shown in green and grey arrows respectively. Reproduced from Silva et al.[32]

The value of simulating isolated proteins has borne out in many subsequent studies. For example, Sun et al. inferred how GPCR-binding to G proteins allosterically triggers nucleotide release from a site over 30 Å away from simulations of G proteins.[35] Porter et al. later used similar logic to identify conformational dynamics in the active sites of isolated myosin motor proteins that are predictive of the life-time of the actin-bound states of myosins.[36] Functionally relevant dynamics have also been observed in simulations of kinases,[37,38] GPCRs,[39] enzymes like SETD8,[40] DNA-binding,[41] and transcription.[42,43]

## Methods and tools for the broader scientific community

Besides the scientific insights, some of Folding@home's greatest legacies are the many algorithms and software that have been developed and shared with the broader scientific community.

MSM methods have been one of the most important contributions.[44-46] As already mentioned, the Pande lab began developing and applying ideas on how to construct and analyze these models in the mid 2000s,[15,16,47] in synergy with concurrent developments in the rest of the field.[48,49] These algorithms were soon incorporated into a software package, called MSMBuilder,[50-52] that was adopted by many theorists for analyzing their simulation data. Later versions of the code drew heavily from another software package developed by the Folding@home community, called MDTraj,[53] that provides many useful functions for operating on and analyzing simulation trajectories. The enspara software[54] also provides extra functionality, especially for dealing with large datasets like those frequently generated on Folding@home and tools for inferring allostery and the effects of mutations.[55,56] Methods have also been introduced for assessing the quality of different MSMs,[57-59] defining states for complex systems,[60] and accounting for memory effects.[61,62] Analysis methods built on the MSM framework facilitate a wide variety of research, from understanding allosteric networks,[55,63,64] to understanding the effects of mutations and drugs,[56] predicting the effect of such perturbations.[65] and reconciling simulations and experiments.[66]

Building off the MSM framework, the Folding@home community has developed a variety of adaptive sampling algorithms for capturing rare events with far less simulation data than brute force simulations would require. The basic idea is to iterate between running simulations, building an MSM, and then using that MSM to decide where to run the next batch of simulations. The first adaptive sampling simulations were geared towards capturing protein folding and aimed to reduce the uncertainty in the slowest process observed in the simulations.[67,68] It was found that adaptive sampling could capture slow processes with far less simulation data than brute force simulations by counteracting simulations' tendency to redundantly repeat the same high-probability events.[69] While these methods were effective at reducing uncertainty in a model, they were less adept at discovering new conformational states. Adaptive seeding methods helped alleviate this issue by using other enhanced sampling methods to quickly explore conformational space, followed by additional simulations to ensure adequate data was gathered from across conformational space to build a good MSM.[70] Starting simulations from poorly sampled states also proved to be a useful way to promote the discovery of new conformational states.[71] However, it was later found that simply choosing poorly sampled states as starting points for new simulations could also lead to pathological outcomes, like spending enormous compute resources exploring high-energy states that were of little relevance compared to higher probability regions of conformational space.[72] Zimmerman et al. introduced the idea of balancing exploration-exploitation trade-offs in a goal-oriented adaptive sampling algorithm called FAST.[73] This algorithm, and related methods like REAP,[74] have proved extremely valuable. The core idea is to balance between preferentially simulating states that optimize some progress variable (e.g. maximizing a distance, or minimizing an RMSD to a target structure) and broad exploration of conformational space. In doing so, the algorithms focus sampling on the relevant regions of conformational space while simultaneously reducing redundant sampling and avoiding dead ends. FAST simulations of systems like 263-residue β-lactamase enzymes capture slow processes with orders of magnitude less simulation time than conventional simulations.[72,73] Furthermore, FAST simulations of the 3,600-residue SARS-CoV-2 spike run on Azure captured enormous conformational changes,[2] while an equivalent simulation time using conventional simulations on supercomputers or Anton2 only captured small fluctuations around the starting conformation.[75] More discussion of the spike can be found in the section on Covid. Here, we simply emphasize that capturing the conformational changes seen with FAST simulations costing about $70K on the cloud would have cost well over $3M with conventional simulations.

The Folding@home community has also contributed software for running individual simulations faster. For example, the team created the first petascale supercomputer by working with Sony to deploy simulation software on the Playstation.[76] They also contributed code that could make use of multi-core CPUs.[17] Moreover, the Pande lab was one of the early groups to recognize that graphics processing units (GPUs) that were developed to quickly update every pixel on a computer monitor could be repurposed to quickly update the position and velocity of every atom in a protein.[77] This led to the development of the widely used OpenMM software.[78]

Finally, Folding@home has made numerous contributions to the evaluation and creation of force fields. Force fields are the parameters that describe interatomic interactions in a simulation. Evaluating their performance is difficult because it requires running enough simulation to explore the relevant conformational space and a means to make quantitative comparisons to experiments. The parallelism of Folding@home makes it well suited for testing many force fields. Multiple papers have developed ways of predicting NMR observables from simulations and comparing the performance of simulations conducted with multiple force fields to experiments.[79,80] Other methods have been developed for modeling spectroscopic

observables.[21,26,81,82] Learning from these comparisons also led the development of new force fields.[83]

## Guiding experiments

An increasing focus of Folding@home, especially since Dr. Greg Bowman became Director in 2018, has been using simulations and MSMs to guide experiments. Making true predictions and then confirming them in subsequent experimental tests is the holy grail of computational biophysics, requiring accurate force fields, sufficient sampling, extraction of useful insights from complex structural ensembles, and a mix of qualitative and quantitative means to predict experimental outcomes.

Early efforts focused on understanding how mutations alter that stability and activity of β-lactamase enzymes.[29] For example, the M182T mutation has long been known to stabilize TEM β-lactamase, but the mechanism remained unclear as different crystal structures of the variant suggested different mechanisms of stabilization. Zimmerman et al. found that M182T stabilizes a helix at a key domain boundary by capping the helix, ultimately stabilizing both domains of the protein.[84] To test their insight, Zimmerman et al. then proposed a number of other mutations to modulate β-lactamase's stability, built MSMs to make quantitative predictions, and then experimentally tested their results. Interestingly, biochemical intuition suggested that an M182N mutations should also stabilize β-lactamase by capping the same helix, but the MSM for this variant predicted the mutation to be neutral due to a competing destabilizing interaction. This and other predictions were confirmed experimentally, showing the power of MSMs to outperform human biochemical intuition. Related work used the same principles to understand how mutations allosterically alter the enzymes activity, inform the design of new variants to modulate activity, and perform experimental tests that confirmed the models' predictions.[85,86]

Folding@home has also been making important contributions to our ability to identify and drug cryptic pockets. These pockets are absent in known crystal structures but form due to protein dynamics. They present a number of new opportunities, from providing a means to target proteins otherwise thought to lack druggable pockets to ways to allosterically enhance desirable activities, a feat that is impossible with traditional drug design given its focus on inhibiting targets by sterically blocking interactions. MSMs of β-lactamase demonstrated that simulations can capture known cryptic pockets. Subsequently, these methods were used to predict novel cryptic pockets in multiple β-lactamases and to design experimental tests of these pockets, which were ultimately confirmed to exist and exert allosteric control over enzyme activity.[64,87] Knoverek et al. then showed that these pockets are functionally relevant, as increasing pocket opening can actually facilitate hydrolysis of some substrates.[86]

Simulations on Folding@home are also helping to improve drug discovery beyond elucidating cryptic pockets. For example, Hart et al. proposed an approach to incorporate protein conformational heterogeneity into rational drug design, called Boltzmann docking and found that it predicts affinities better than docking against single crystal structures.[85] The algorithm works by docking compounds against representative conformations of each state of an MSM and then ranking them based on a population-weighted average docking score. In addition to naturally considering a set of structures rather than having to pick a single structure from a large simulated ensemble, the method also naturally favors compounds that bind to higher probability conformations and therefore have to pay a lower penalty to stabilize their target conformation. Hart et al. then used this approach to design and experimentally confirm novel β-lactamase inhibitors that target a cryptic pocket found by simulation.[88] Meller et al. later derived a more physically-rigorous version of this algorithm, called MSM-docking, and used it to guide experimental predictions of binding affinities and experimental tests.[89]

Now these methods are being applied broadly to a number of biologically important proteins. For example, Cruz et al. predicted a cryptic pocket in an Ebola protein that was thought to lack druggable pockets based on known crystal structures, and then experimentally confirmed the existence of this pocket and its allosteric control over a key RNA-binding interaction (Fig. 4).[90] Meller et al. used these tools to understand how sequence variations in myosin motors control the probability of pocket opening and, thereby, modulate the affinity for the allosteric inhibitor blebbistatin.[89] To test this approach, they predicted the affinity of blebbistatin for a new motor and experimentally confirmed their prediction. MSMs are also being used to understand how mutations modulate the structural preferences of the ApoE protein and, ultimately, people's risk of developing Alzheimer's Disease.[26]

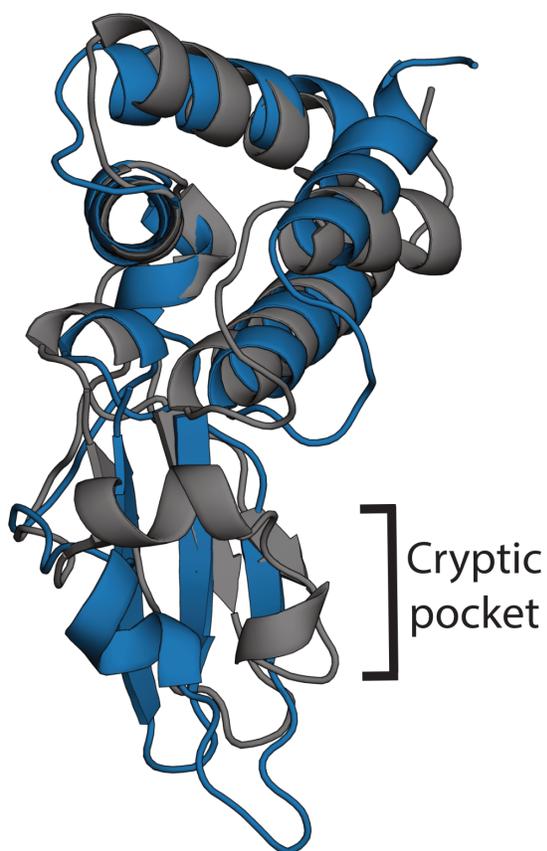

Fig. 4. A structure with a cryptic pocket in Ebola's VP35 protein (blue) overlaid with the crystal structure (gray). The existence of this pocket and its allosteric control over RNA binding were confirmed using chemical labeling experiments. Reproduced from Cruz et al.[90]

## COVID-19

Folding@home's strengths and past achievements positioned it to pivot quickly to address the COVID-19 pandemic. In February 2020, just before lockdowns started in the United States, the Chodera lab launched the first Folding@home simulations of SARS-CoV-2 viral proteins based on emerging structural data.[91] The community's response was literally overwhelming: over the next month, Folding@home's user base grew from ~30K active users to over one million,[2] causing server-side infrastructure to stagger under the load. This crisis was ameliorated with an outpouring of generosity from numerous tech organizations, who offered help scaling up Folding@home's infrastructure in the cloud. At its peak, the community's aggregate compute power was estimated to be at least 5-fold greater than the Summit supercomputer, which was the world's fastest traditional supercomputer at the time.

A key goal during this time was to contribute to the development of new therapeutics by exploiting Folding@home's ability to computationally test many possible drugs in parallel. Towards this end, Folding@home joined in in an international collaboration—the COVID Moonshot [https://covid.postera.ai/covid]—to develop a patent-free small molecule protease inhibitor that would serve as an inexpensive therapy, as well as a prophylactic for high-risk individuals.[92] Following high-throughput crystallization studies by the UK's Diamond Light Source that identified multiple molecules that bind the SARS-CoV-2 main protease,[93] the COVID Moonshot project was formed to accelerate the development of potent inhibitors for clinical trials. Folding@home performed absolute and relative free energy calculations[94] on a massive scale to prioritize new molecules for synthesis and subsequent experimental tests. In principle, these calculations—which sample protein and ligand motions—are far more accurate than computational docking, but enormously expensive, generally limiting these calculations to a few dozen at a time. Using the parallelism of Folding@home, the team quickly screened tens of thousands of potential inhibitors. Many iterations of computational design and testing followed, leading to the development of novel chemical scaffolds that are now progressing towards clinical trials.[95,96]

Folding@home's parallelism also enabled the scientific team to search through the viral proteome for new potential drug targets. Indeed, cryptic pockets were found across much of the SARS-CoV-2 proteome. A noteworthy example is a cryptic pocket in the protein Nsp16 (Fig. 5).[97] This pocket is shared across extant coronaviruses but is absent in human homologs of the protein. Therefore, drugs that target this site could inhibit all coronaviruses without impacting similar human proteins and causing unwanted side-effects.

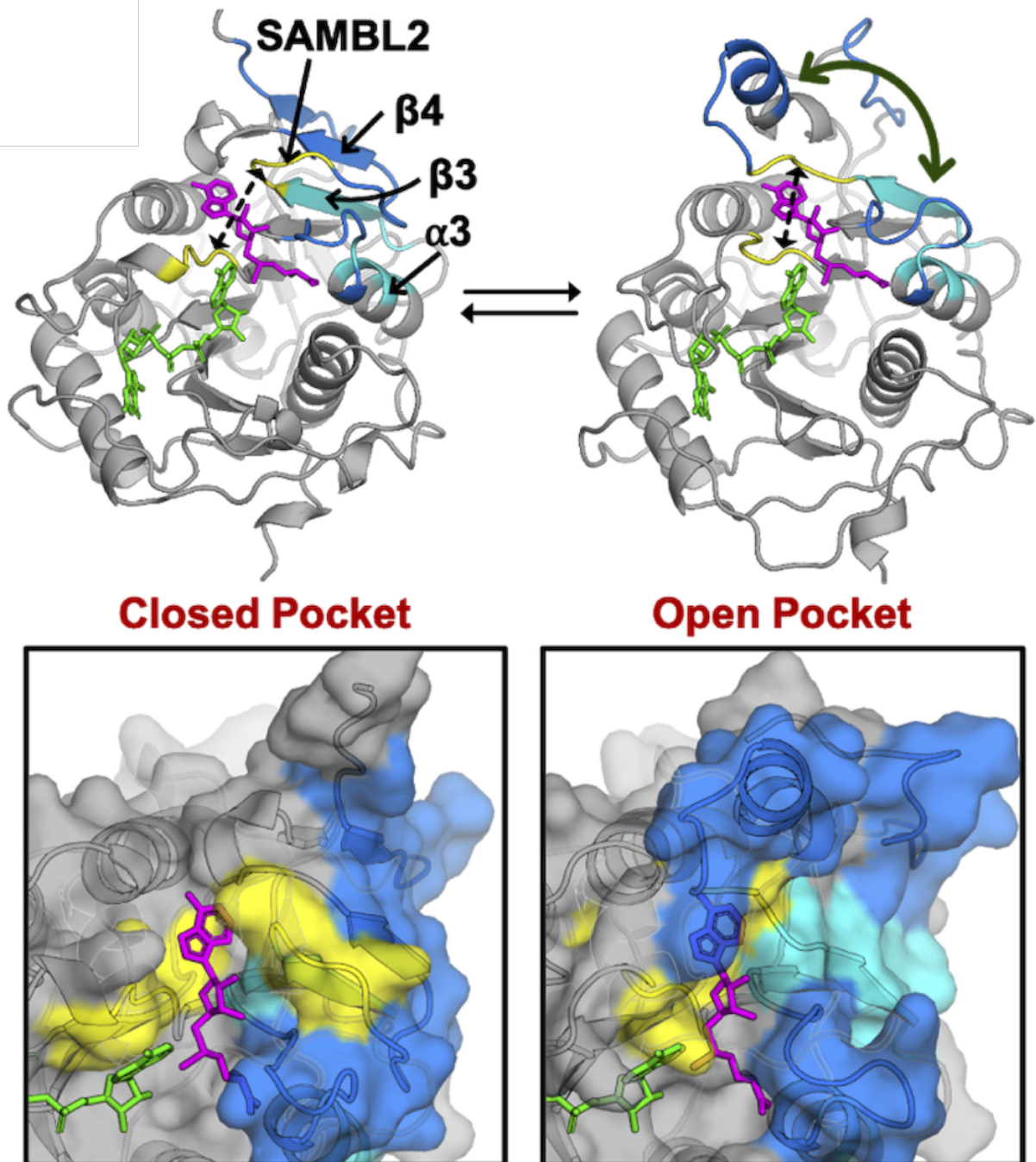

Fig. 5. Structural states with the Nsp16 cryptic pocket closed and open, showing how pocket opening is correlated with collapse of the active site's SAM-binding pocket. The insets show surface views of the closed and open pocket. Residues exposed upon pocket opening are shown in cyan and the regions undergoing the opening motion are shown in blue. S-adenosyl-L-Methionine (SAM) is in magenta sticks, and the RNA substrate is in green sticks. Collapse of the SAM-binding pocket is measured as the distance between two loops labeled SAMBL2 and gate loop 2, shown in yellow. Reproduced from Vithani et al.[97]

Protein variants from different coronaviruses were also compared to understand what factors make the SARS-CoV-2 virus so infectious and point towards new ways to thwart the virus. For example, spike proteins from multiple coronaviruses were simulated to understand how differences in their conformational preferences alter the viruses' ability to enter cells and evade host immunity. MSMs built from these simulations revealed that the spike undergoes a dramatic opening motion that is far larger than was expected from cryoEM data and that controls a tradeoff between cell entry and immune evasion (Fig. 6).[2] Viruses whose spike proteins spend more time in the open state are better at cell entry, but they are also more susceptible to recognition by our immune systems. One factor that enabled SARS-CoV-2 to become a pandemic is that its spike is more closed, making it better at immune evasion. However, the virus maintains its ability to enter cells by acquiring mutations that increase the affinity between an open spike and its binding partner on host cells, angiotensin-converting enzyme 2 (ACE2). The result is that SARS-CoV-2 is better at evading our immune systems than the original SARS virus and just as good at cell entry. Subsequent viral variants have optimized these factors further.

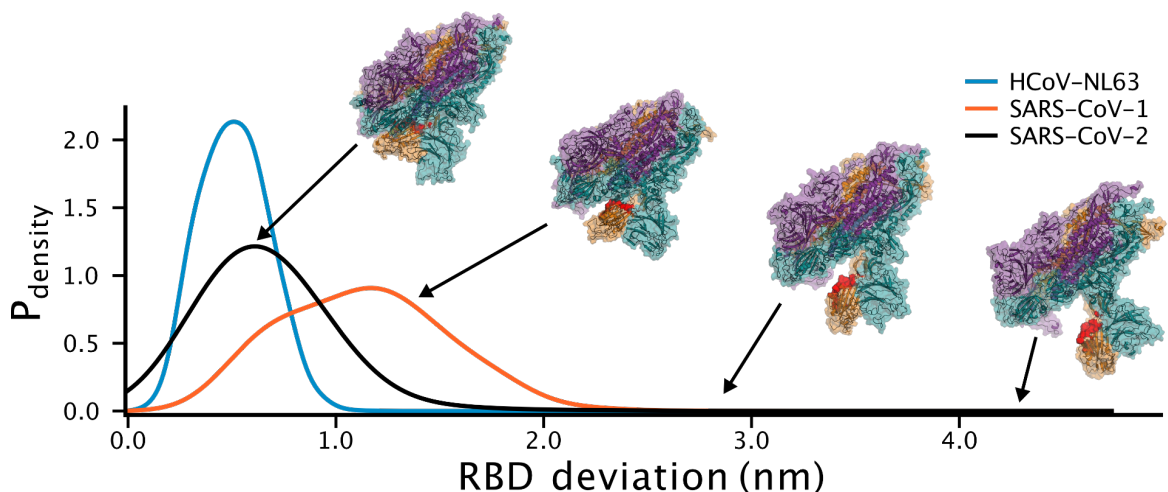

Fig. 6. The probability distribution of spike opening for three spike homologues. Opening is quantified in terms of how far the center of mass of a receptor-binding domain (RBD) deviates from its position in the closed (or down) state. The cryptic epitope for the antibody CR3022 (red) and the ACE2-binding interface are both exposed in open structures but buried in closed structures. As a result, more open spikes are better at cell entry but more susceptible to host immunity. Reproduced from Zimmerman et al.[2]

Other simulations interrogated how existing NTP-analogs (like remdesivir) target viral RNA replication,[98] the function of the SARS-CoV-2 envelope protein (an ion channel and membrane protein crucial to viral assembly),[99] and the design of ACE2 decoys that could block the spike from engaging host ACE2.[100]

The pandemic also exemplifies the ways that Folding@home has impact beyond its direct contributions to science. For example, many volunteers reported mental health benefits from the opportunity to take action at a time when they otherwise felt helpless. Others continue to take a keener interest in science, with outcomes ranging from continued involvement in Folding@home to pursuing STEM careers. The pandemic also accelerated trends in open science. All the COVID-19 data and results generated by Folding@home are publicly available through the COVID-19 Structure and Therapeutics Hub developed jointly with the Molecular Sciences Software Institute [http://covid.molssi.org] to empower other researchers, and this practice is being continued as the scientific team returns to working on other problems.

## Outlook

Folding@home has enabled numerous scientific insights by steadily providing the ability to simulate longer timescales, larger systems, and more numerous variants. Work on SARS-CoV-2 exemplifies the ways that many different proteins can be studied in parallel to find the best drug targets, many variants of a protein can be compared to understand how they function, and many small molecules can be screened to find the best potential drugs. We are already seeing these use patterns continue into the future. The AI-driven Structure-enabled Antiviral Platform (ASAP) is working to automate the discovery of oral antivirals for pandemic preparedness. Comparative studies of large systems like myosin motors are providing new insight into sequence-ensemble-function relationships.[36,89] Ongoing work screening the folding and binding properties of peptides and peptidomimetics on a massively parallel scale[101-103] may help identify new therapeutics. And the same platform is being used to study many other systems.

Folding@home has given researchers compute capability a decade ahead of their peers, leading to the creation of novel computing paradigms (e.g. GPU computing with OpenMM) and algorithms for today's cloud-based compute (e.g. MSMs).  Looking ahead to the exascale era and beyond, Folding@home is poised to be an essential tool for research, serving as a general platform for addressing pressing problems in both basic science and human disease, while remaining ready to rapidly respond to emergent health threats. Citizen-scientists, unite!

To participate, visit https://foldingathome.org/start-folding/.

## Acknowledgements

Thanks to the entire Folding@home community for all their contributions. G.R.B. was funded by NSF MCB 2218156, NIH R01GM124007, NIH AG067194, and a Packard Fellowship. V.A.V. is supported by NIH R01GM123296.  Thanks to John Chodera for helpful comments.

## Disclosures

G.R.B. is a co-founder and board member of Decrypt Biomedicine.

## Author contributions

VAV, VSP, and GRB wrote the paper.